\documentclass[prb,aps,twocolumn,showpacs]{revtex4-2}

\usepackage{graphicx,color}
\usepackage{amsthm}
\usepackage{amsfonts}
\usepackage{algorithmic}
\usepackage{enumerate}
\usepackage{latexsym}
\usepackage{amsmath}
\usepackage{amssymb}
\usepackage{bm}
\usepackage[pdftex,plainpages=false,colorlinks=true,linkcolor=blue, citecolor=blue, urlcolor=blue]{hyperref}

\emergencystretch=\maxdimen
\begin{document}
\title{The odd-parity altermagnetism: A spin group study}

\author{Minghuan Zeng$^{1}$}
\author{Zheng Qin$^{1}$}
\author{Ling Qin$^{2}$}
\author{Shiping Feng$^{3,4}$}
\author{Lin Wu$^{5}$}
\email{wulin@cqu.edu.cn}
\author{Dong-Hui Xu$^{1,6}$}
\email{donghuixu@cqu.edu.cn}
\author{Rui Wang$^{1,6}$}
\email{rcwang@cqu.edu.cn}

\affiliation{$^{1}$Institute for Structure and Function \& Department of Physics \& Chongqing Key Laboratory for Strongly Coupled Physics, Chongqing University, Chongqing, 400044, P. R. China}

\affiliation{$^{2}$College of Physics and Engineering, Chengdu Normal University, Chengdu, 611130, Sichuan, China}

\affiliation{$^{3}$Department of Physics, Faculty of Arts and Science, Beijing Normal University, Zhuhai, 519087, China}

\affiliation{$^{4}$School of Physics and Astronomy, Beijing Normal University, Beijing, 100875, China}

\affiliation{$^{5}$College of Materials Science and Engineering, Chongqing University, Chongqing, 400044, China}

\affiliation{$^{6}$Center of Quantum materials and devices, Chongqing University, Chongqing 400044, P. R. China}

\begin{abstract}
Following recent intensive studies on altermagnetism(ALM) characterized by non-relativistic even-parity spin splitting,
realizing unconventional odd-parity magnetism has also attracted increasing interest.
Here, using symmetry arguments based on spin-group analyses, we elucidate sufficient conditions
for the emergence of odd-parity spin splitting in collinear antiferromagnetic systems,
which is further established as the standard odd-parity ALM. It is derived that the odd-parity ALM arises from the following criteria:
(i)the breaking nonmagnetic time reversal symmetry(TRS), {\color{blue}i.e., the breaking real-space TRS}; (ii)the long-range collinear compensated magnetism;
(iii)the symmetry $[C_{2}||\bar{E}]$ or $[C_{2}||M]$ connecting opposite-spin sublattices,
where $C_{2}$, $\bar{E}$, and $M$ respectively represent a $180^{\circ}$ rotation around the axis perpendicular to spins, the inversion,
and the mirror reflection separating opposite-spin sublattices, directly reflecting the high-order harmonic($l\ge3$) and the $p$-wave($l=1$) odd-parity ALM, respectively.
Moreover, we utilize the well-known Haldane-Hubbard model to identify odd-parity spin splitting in the collinear ALM ground state,
where (i)the nonmagnetic TRS is broken by opposite sublattice currents coming from the Haldane hopping;
(ii)the symmetry $[C_{2}||\bar{E}]$ is ensured because the currents flowing on opposite-spin sublattices are reversed.
\end{abstract}

\maketitle


{\it Introduction.}---Recent series of theoretical and experimental reports\cite{Moreno12,Noda16,Okugawa18, Ahn19,Hayami19,Naka19,Libor20,Yuan20,Hayami20, Mazin21,Yuan21,Naka21,Rafael21,Shao21,Libor22,Bai22,Bose22,Karube22} have driven attention to time-reversal-symmetry(TRS)-breaking even-parity altermagnetism(ALM)
in materials with collinear-compensated magnetic order, which is incompatible with conventional ferromagnetism(FM)\cite{Landau65} or antiferromagnetism(AFM)\cite{Brinkman66,Neel71,Corticelli22}.
The even-parity ALM has been shown as the subset of the crystal symmetry paired spin-momentum locking in AFM systems\cite{MaHY22,HuM25}.
The alternating spin polarizations in both the real and momentum space characteristic of  this unconventional magnetic phase\cite{Libor22_1,Libor22}
give rise to multiple TRS breaking responses
such as the anomalous Hall effect\cite{Libor20,Feng22}, charge to spin conversion\cite{Rafael21}, spin-splitter torque\cite{Rafael21,Karube22},
and giant tunneling magnetoresistance effects\cite{Libor22,Shao21}.
By employing and developing the spin-group formalism\cite{Brinkman66,Litvin74,Litvin77} of symmetry transformations
in the decoupled real and spin space, researchers have established principles to look for materials having the TRS breaking even-parity ALM.
It is revealed clearly that the energy scale and momentum dependence of spin splitting are
determined by the anisotropic electric crystal potential\cite{Libor22_1,Libor22_2,Karube22,Bose22,Bai22}.

\begin{figure}
\includegraphics[scale=0.27]{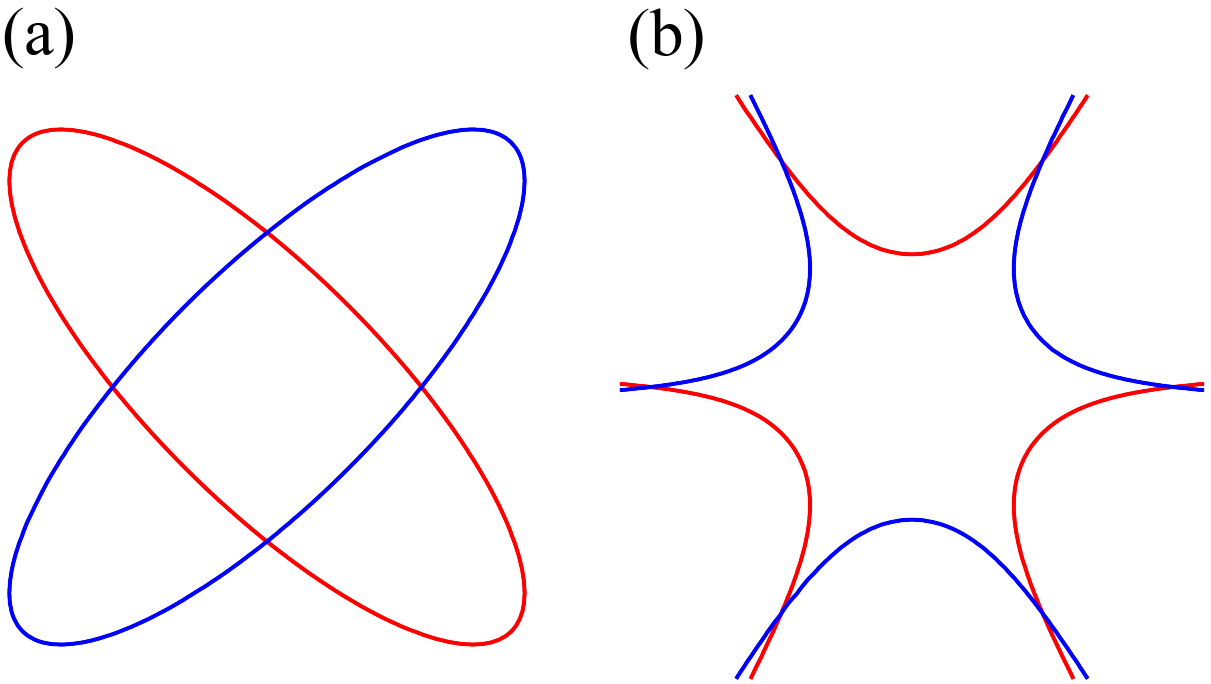}
\caption{(Color online) The schematic cartoon of the spin splitting in momentum space for (a)even- and (b)odd-parity altermagnets with $d$- and $f$-wave
spin splitting. Here the red and blue solid lines represent the up- and down-spin electron energy dispersions, respectively.
\label{Illust-ALM}}
\end{figure}

In previous studies, nonmagnetic crystal structures are assumed time-reversal-symmetric,
which naturally leads to that ALM induced by the anisotropic electric crystal potential is of even parity\cite{Libor22_1,Libor22_2,Bai24}.
The non-relativistic odd-parity magnetism has been heatedly discussed recently and become a rapidly developing research field\cite{Hellenes24,
Zhu25,Liu25,Song25,Brekke24,Ezawa25,Yu25,Lin25}.
The odd-parity magnetism has TRS spin-momentum locking analogous to the Rashba and
Dresselhaus spin-orbital coupling, implying its promising application in spintronics\cite{Leggett75,Zutic04,Zhang14,Koo20}.
However, odd-parity spin splitting observed up to now has been confined to
coplanar spin configurations\cite{Hellenes24,Song25,Yu25}, in strong
contrast to even-parity spin splitting in collinear altermagnets\cite{Libor22_1,Libor22_2,Bai24}.
While distinguishing and manipulating the coplanar magnetism are not easy experimentally,
realizing the odd-parity collinear spin-splitting is now at the core of recent studies on magnetism.
{\color{blue}On the other hand, several works have revealed that the light\cite{Zhu25,Liu25}, sublattice current\cite{Lin25},
and orbital order\cite{Zhuang25} are efficient to enable the occurrence of odd-parity ALM,
which has the distinct spin splitting from even-parity altermagnets.
As schematically illustrated in Fig.\ref{Illust-ALM}(a) and (b), there appears significant spin splitting in momentum space
for both even- and odd-parity ALM's, while their form factors are quite different.
Specifically, for altermagnets having the symmetry $[C_2||C_{nz}^{1}]$ where $C_2$ and $C_{nz}^{1}$ represent a $180^\circ$ rotation
around the axis perpendicular to spins and a real-space rotation of $\tfrac{2\pi}{n}$ radian around $z$ axis, respectively,
acting this operation on the spin- and crystal-momentum-dependent bands $E_{\bm{k}\sigma}$ for $n/2$ times gives rise to
\begin{equation}\label{Symmetry-Constraint}
[C_2||C_{nz}^{1}]^{n/2}E_{\bm{k}\sigma} =
\left\{ \begin{array}{ll}
E_{-\bm{k}\sigma},&, n/2 \text{ is even} \\
E_{-\bm{k}-\sigma},&, n/2 \text{ is odd}
\end{array} \right.\;.
\end{equation}
Therefore, for even(odd)-parity ALM's, the energy bands with spin splitting in momentum space are subjected to the relation $E_{\bm{k}\sigma}=E_{-\bm{k}+(-)\sigma}$.
Meanwhile, opposite-spin sublattices in even- and odd-parity ALM's cannot be connected by the operation $[C_2||\bm{\tau}]$
with $\bm{\tau}$ being the real-space translation between them, while is connected by $[C_2||C_{nz}^{1}]$ with $n/2$ being even and odd, respectively.}

Previous works\cite{Lin25,Zhu25,Liu25} have realized the odd-parity ALM by breaking {\color{blue}the nonmagnetic TRS that is the TRS in real space\cite{Sun08}},
while there lacks a systematic investigation based on the spin-group formalism\cite{Brinkman66,Litvin74,Litvin77}
to unveil the underlying relationship between the breaking  nonmagnetic TRS and the occurrence of odd-parity ALM.
In this Letter, we first adopt the spin group method\cite{Brinkman66,Litvin74,Litvin77} to derive sufficient conditions for the occurrence of odd-parity ALM
that are closely associated with the breaking nonmagnetic TRS. Then Haldane-Hubbard model is studied using the cluster slave spin method\cite{WCLee17,Zeng21,Zeng22}
to reveal the sublattice-current-induced odd-parity spin splitting in the ALM ground state.

{\it Symmetry analysis.}---The theory of spin groups with the element $[R_i||R_j\bm{v}]$ is established to describe the symmetry of spin configurations in crystals\cite{Litvin74,Litvin77,Libor22_1}. In an element $[R_i||R_j\bm{v}]$, $R_i$ and $R_j$ represent respectively the proper or improper rotation matrices
in spin and real space, and $\bm{v}$, a column matrix, represents the real-space translation. In this sense, applying a spin group element $[R_i||R_j\bm{v}]$
to the spin arrangement $\bm{S}(\bm{r})$ gives rise to\cite{Litvin77}
\begin{equation}\label{Group-Elemts}
[R_i||R_j\bm{v}]\bm{S}^{\alpha}(\bm{r})=\sum_{\beta=1}^{3}R_{i\alpha\beta}\bm{S}^{\beta}([R_j\bm{v}]^{-1}\bm{r})\;,
\end{equation}
{\color{blue} where $\bm{S}(\bm{r})=(S^{x}(\bm{r}),S^{y}(\bm{r}),S^{z}(\bm{r}))$ is a vector function of the position $\bm{r}$.
Therefore the left side of Eq.\eqref{Group-Elemts} represents the operation under the active viewpoint in group theory,
whereas its right side represents the spin-group operation under the passive viewpoint.}

Spin groups can be decomposed as a direct product $\bm{r}_s \times \bm{R}_s$ of the spin-only group $\bm{r}_s$
and the nontrivial spin groups $\bm{R}_s$ containing the elements $[R_i||R_j]$\cite{Litvin74,Litvin77,Libor22_1}, where the translation vector $\bm{v}$
is abridged because we focus on discussing the spin splitting in momentum space. A nontrivial spin group $\bm{R}_{s}$
belongs to the family of $\bm{B}$ and $\bm{R}$ if the components on the left of its group element constitute the group of $\bm{B}$,
while the right-hand components the group of $\bm{R}$. The nontrivial spin group contains a normal subgroup $[E||\bm{R}]$,
and its right-hand components constitute a normal subgroup $\bm{r}$ of $\bm{R}$. Moreover, the factor group $\bm{R}/\bm{r}$ is isomorphic with $\bm{B}$\cite{Litvin77}.
In this sense, the group $\bm{R}$ and $\bm{B}$ can be expressed into the coset decomposition with respect to the normal subgroup $\bm{r}$
\begin{equation}
\bm{R} = \bm{r} + R_2 \bm{r} + \cdots + R_n \bm{r} \;,
\end{equation}
and
\begin{equation}
\bm{B} = E + B_2 + \cdots + B_n \;.
\end{equation}
Then the nontrivial spin group $\bm{R}_s$ can be constructed by pairing the group element $B_i$ and $R_i$
in terms of one isomorphic mapping between them, i.e.,
\begin{equation}
\bm{R}_{s} = [E||\bm{r}] + [B_2||R_2][E||\bm{r}] + \cdots + [B_n||R_n][E||\bm{r}]\;.
\end{equation}

For collinear magnets, the spin-only group reads $\bm{r}_s = \{ C_{\infty},\bar{C}_2 \}$\cite{Litvin74,Litvin77,Libor22_1},
where $C_{\infty}$ is a group consisting of all rotations around the common axis of spins; $\bar{C}_2$ represents a 180$^{\circ}$ rotation
around the axis perpendicular to spins, followed by the spin-space inversion. For the magnetic system with nonmagnetic TRS, the
symmetry $\bar{C}_2$ is always accompanied by the time reversal in real space, such that $[\bar{C}_2||T]$ is the symmetry of these magnetic systems
with $T$ only acting in real space, enforcing which on the electron energy dispersion gives rise to,
\begin{equation}
[\bar{C}_2||T]E_{\bm{k}\sigma} = E_{-\bm{k}\sigma}\;.
\end{equation}
Therefore the symmetry $[\bar{C}_2||T]$ ensures the inversion symmetry of the electron energy dispersion, i.e., $E_{\bm{k}\sigma}=E_{-\bm{k}\sigma}$,
independent of whether the inversion symmetry {\color{blue} is present} in the crystallographic Laue group or not\cite{Libor22_1,Libor22_2}.

The spin component of the nontrivial spin group $\bm{R}_s$ is chosen as $\bm{B}=\{ E \}$ or $\{ E,C_2 \}$ in collinear
magnets\cite{Libor22_1,Libor22_2}. In this sense, the nontrivial spin group can be constructed in 3 different manners,
which corresponds to three distinct magnetic states:
(i) for $\bm{B}=\{ E \}$, there exists a single coset decomposition, i.e., $\bm{B}=\{ E \}$. Then with the crystallographic Laue group $\bm{G}$,
the nontrivial spin group is written as $\bm{R}_{s}^{I} = \{E||\bm{G}\}$, which describes the universal magnetization in ferromagnets;
(ii) for $\bm{B}=\{ E,C_2 \}$, the one-coset decomposition becomes $\bm{B}$ itself, and then the nontrivial spin group is expressed as
$\bm{R}_{s}^{II} = \{E||\bm{G}\}+\{C_2||\bm{G}\}$, directly describing the spin symmetry of collinear AFM's;
(iii) for $\bm{B}=\{ E,C_2 \}$, the other coset decomposition reads $\bm{B}=\{ E \} + C_2\{ E \}$.
The two-coset decomposition of the crystallographic Laue group $\bm{G}$ is written as $\bm{G} = \bm{H} + A\bm{H}$
{\color{blue}with $A$ and $\bm{H}$ being the real-space proper or improper rotation and its halved subgroup, respectively}.
Then the nontrivial spin group is constructed as $\bm{R}_s^{III} = \{E||\bm{H}\} + \{C_2||A\bm{H}\} =  \{E||\bm{H}\} + \{C_2||\bm{G}-\bm{H}\}$,
describing the symmetry of the spin arrangement in even-parity altermagnets. The elements in $\{C_2||\bm{G}-\bm{H}\}$
determine spin polarizations in real and momentum space, and ensure the vanishing net magnetization in altermagnets.
{\color{blue} It is worth emphasizing that $A$ cannot be the inversion $\bar{E}$, because on account of which the symmetry $[C_2||E] = [\bar{C}_2||T][C_2||\bar{E}]$ emerges,
and thus the halved subgroup $\bm{H}$ and its coset $A\bm{H}$ have the elements $E$, $A$, and $A^{-1}$ in common,
therefore we have $\bm{H}=A\bm{H}=\bm{G}$. In this sense, for systems with long-range compensated collinear magnetism,
the presence of the symmetry $[C_2||\bar{E}]$ enables its spin group $\bm{R}_{s}$ to belong to type II that
describes the symmetry of spin arrangement in conventional antiferromagnets.}

{\color{blue}
However, for collinear antiferromagnets with the breaking nonmagnetic TRS that can be caused by sublattice currents\cite{Lin25}, light\cite{Liu25,Zhu25},
or the phase-dependent hopping coming from orbital orders\cite{Zhuang25}, the symmetry $[\bar{C}_2||T]$ appearing in even-parity ALM's no longer holds.
In contrast, the identity element in the nontrivial spin group $\bm{R}_{s}$ for collinear magnets, $[\bar{C}_2||E]$, remains with $E$ being the identity in real space.
Furthermore, as shown by Eq.\eqref{Symmetry-Constraint}, the odd-parity spin splitting in collinear magnets is ensured by the symmetry $[C_2||C_{2z}]$,
while for two-dimension systems, the operation $C_{2z}$ is equivalent to $\bar{E}$,
such that the odd-parity spin splitting in two-dimension magnetic systems with long-range compensated collinear magnetism is
ensured by the symmetry $[C_2||\bar{E}]$.
In addition, for $p$-wave altermagnets with spin splitting perpendicular to the mirror reflection $M$, which indicates that $[C_2||M]E_{\bm{k}\sigma}=E_{M\bm{k}-\sigma}=E_{\bm{k}\sigma}$, the $p$-wave spin splitting is ensured by the symmetry $[C_2||M]$.
Therefore, in strong contrast to even-parity ALM's, the long-range compensated collinear magnets exhibiting odd-parity
spin splitting have the following characteristics:
(i)the breaking nonmagnetic TRS then the breaking inversion symmetry of the Bravais lattice, i.e., $\bar{E}_{BL}$;
(ii)the symmetry $[C_2||\bar{E}]$ or $[C_2||M]$, which directly enables that the nontrivial spin group expressed as
$\bm{R}_s^{IV} = \{E||\bm{H}\} + \{C_2||A\bm{H}\} =  \{E||\bm{H}\} + \{C_2||\bm{G}-\bm{H}\}$ has the distinct
coset $A\bm{H}$ of the halved crystallographic Laue group $\bm{H}$ containing the inversion $\bar{E}$ or the mirror reflection $M$.
To sum up, in spit of the similar form of the nontrivial spin group to even-parity ALM's, the odd-parity spin splitting in compensated collinear magnets
need to be classified into a new type, i.e., odd-parity altermagnetism, not only because the symmetry $[\bar{C}_2||T]$ is absent,
but also the inversion symmetry $\bar{E}$ or the mirror reflection $M$ is contained in the coset $A\bm{H}$. }

\begin{figure}
\includegraphics[scale=0.35]{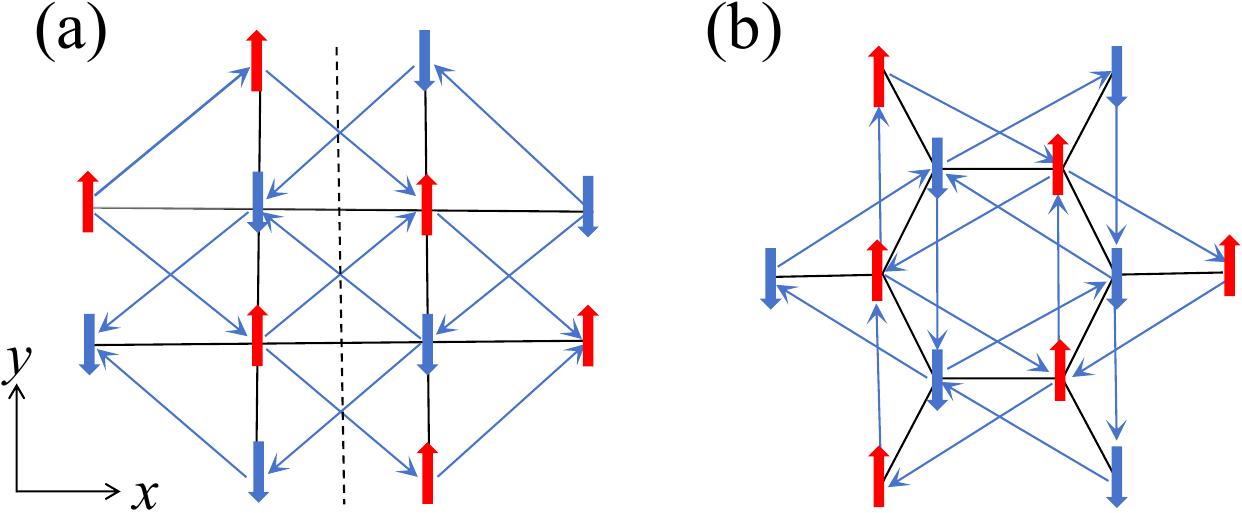}
\caption{(Color online) The schematic illustration of the (a) square- and (b) hexagon-lattice compensated collinear magnets
with opposite sublattice currents denoted by blue solid arrows, where the up-(red) and downward(blue) solid arrows represent the corresponding spin polarizations.
Here the black dashed line denotes the mirror reflection $M_{yz}$ perpendicular to $x$ axis.
\label{sub-Current}}
\end{figure}
{\it Haldane-Hubbard model.}---{\color{blue}We use Haldane-Hubbard(HH) model\cite{Haldane88} as an example
where the nonmagnetic TRS is broken by sublattice currents to verify sufficient conditions for the occurrence of odd-parity ALM.
As shown in Figs.\ref{sub-Current}(a) and (b), the nonmagnetic TRS is broken by opposite currents flowing on two sublattices
in the square- and hexagon-lattice magnetic system. In the square-lattice system[cf. Fig.\ref{sub-Current}(a)],
the reversed current flowing on opposite-spin sublattices
enables the reversed chirality between them, then the compensated collinear magnetism triggers the emergence of a new form of
symmetry $[C_2||M_{yz}]$ with $M_{yz}$ being the mirror reflection perpendicular to $x$ axis as denoted by the black dashed line,
which then gives rise to the $p$-wave ALM\cite{Lin25}. In addition, the reversed sublattice currents enables a new form of symmetry
$[C_2||\bar{E}_{BL}\bm{\tau}]$ with $\bm{\tau}=\hat{x}, \hat{y}$ being the minimal vector connecting opposite-spin sublattices, indicating that this
$p$-wave altermagnet has the time-reversal-symmetric energy band, i.e., $E_{\bm{k}\sigma}=E_{-\bm{k}-\sigma}$, in strong contrast to conventional $p$-wave magnets\cite{Hellenes24,Song25,Brekke24}. Therefore, the nontrivial spin group $\tilde{\bm{R}_{s}}$ in the square-lattice system can be expressed as
$\tilde{\bm{R}}_{s}-[C_2||\bar{E}_{BL}\bm{\tau}] = [E||\bm{H}]+[C_2||M_{yz}\bm{H}]$ and $\bm{H}=\{E\}$.
However, as shown in Fig.\ref{sub-Current}(b), sublattice currents in the hexagon-lattice system have the same chirality between opposite-spin sublattices,
thus the symmetry $[C_2||C_{6z}]$ appears as the staggered magnetization sets in with $C_{6z}$ being the rotation of $\pi/3$ radian around the axis
perpendicular to the plane. In addition, opposite-spin sublattices with the reversed current are connected by the symmetry $[C_2||\bar{E}_{BL}\bm{\tau}]$
with $\bm{\tau}$ being the minimal vector connecting them. Therefore, the nontrivial spin group can be written as $\tilde{\bm{R}}_{s}-[C_2||\bar{E}_{BL}\bm{\tau}] = [E||\bm{H}]+[C_2||C_{6z}\bm{H}]$ and $\bm{H}=\{E, C_{6z}^{2}, C_{6z}^{4}\}$, directly describing a $f$-wave spin splitting in real and momentum space.

To sum up, the nontrivial spin group for odd-parity ALM's can be expressed as $\tilde{\bm{R}}_{s}=\bm{R}_{s}\oplus \{ [\bar{C}_2||E], [C_2||\bar{E}_{BL}\bm{\tau}] \}$
with $\bar{E}_{BL}$ and $\bm{\tau}$ respectively being the inversion of the Bravais lattice and the minimal vector connecting opposite-spin sublattices,
while the subgroup $\bm{R}_{s}$ reads $\bm{R}_s = \{E||\bm{H}\} + \{C_2||A\bm{H}\} =  \{E||\bm{H}\} + \{C_2||\bm{G}-\bm{H}\}$
with $A$ being the inversion $\bar{E}$ or the mirror reflection $M$, respectively.
Therefore, we have systematically studied the nontrivial spin group in the square- and hexagon-lattice HH model[cf. Fig\ref{sub-Current}],
where sufficient conditions for the odd-parity ALM are satisfied: (i)the nonmagnetic TRS is broken by opposite sublattice currents;
(ii)the nontrivial spin subgroup reads $\bm{R}_s = \{E||\bm{H}\} + \{C_2||A\bm{H}\}$ with $A$ being the inversion $\bar{E}$ and the mirror reflection $M_{yz}$ in the
hexagon- and square-lattice system, respectively.
}

\begin{figure*}
\includegraphics[scale=0.34]{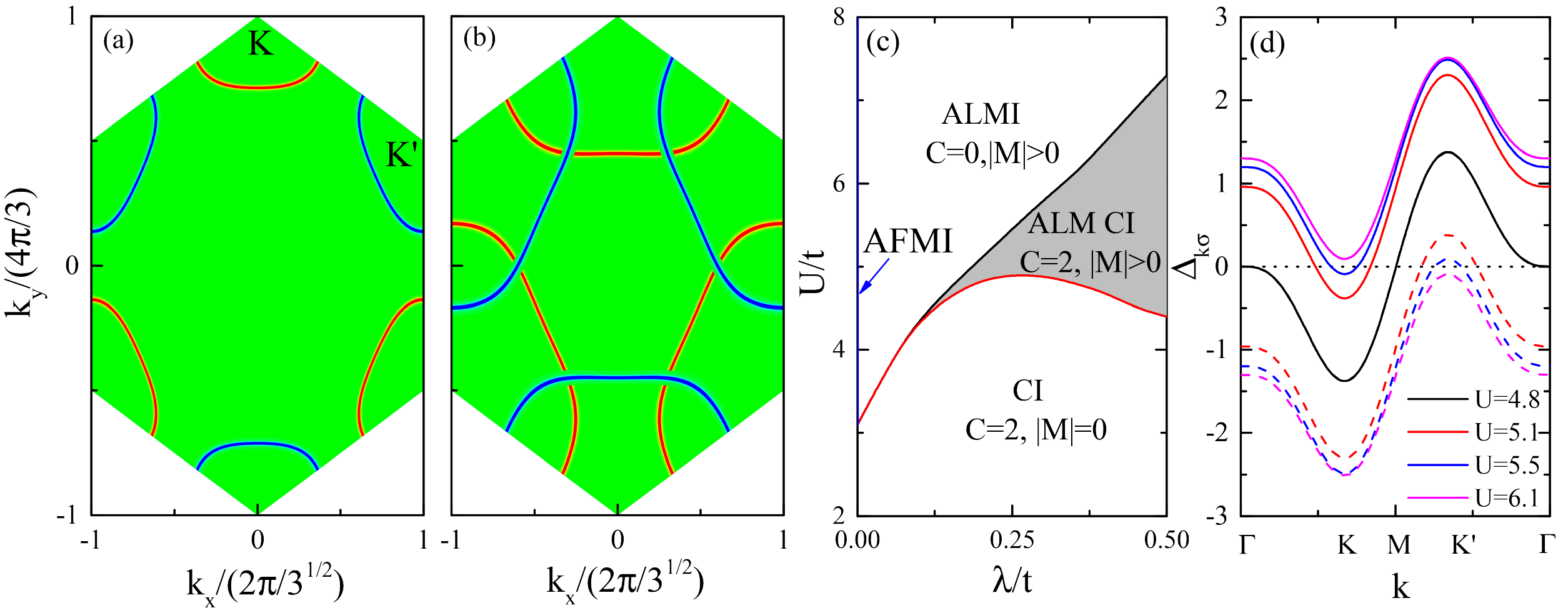}
\caption{(Color online) The electron spectrum function $A_{\sigma}(\bm{k},\omega)$ as a function of momentum at $\lambda=0.3$ and $U=5.5$
in the reduced first Brillouin zone with $\omega=$ 1(a) and 2(b), respectively, where the CCE for up- and down-spin electrons are plotted respectively in red and blue color,
and $\bm{K}^{(')}$ denotes two inequivalent Dirac points;
(c)The phase diagram of the half-filled HH model as a function of Haldane hopping $\lambda$ and onsite electron
Coulomb repulsion $U$, which is composed of four regimes: AFMI at $\lambda=0$ for $U>U_{\rm AFM}\approx 3.1$, CI with the TKNN number $C=2$ at small $U's$,
the odd-parity ALM CI with $C=2$ and $|M|>0$ for intermediate interaction strengths, and odd-parity ALMI
in the strong interaction regime. (d)The momentum dependence of the energy gap $\Delta_{\bm{k}\sigma}$[See Eq.\eqref{Gap-Function}]
along the zigzag direction in the first Brillouin zone[See Fig.1 in the Sup. Mat.\cite{Zeng25-supp}] with $\lambda=0.3$ as well as $U=$ 4.8(black), 5.1(red),
5.5(blue), and 6.1(magenta), where $\Delta_{\bm{k}\sigma}$ for up- and down-spin electron quasiparticles are plotted by solid and dashed lines, respectively.
\label{E-EDisp-PD_GAP}}
\end{figure*}

Bellow, HH model is restudied to further reveal the odd-parity spin splitting in momentum space which reads
\begin{eqnarray}\label{HH-Model}
{\rm H} &=&\sum_{\langle ij\rangle}\big[ -tC_{iA}^{\dagger}C_{jB}+\mathrm{H.C.} + \mu\delta_{ij} \sum_{s}C_{is}^{\dagger}C_{js} \big] \nonumber\\
 &+& \lambda\sum_{s=A,B}\sum_{\langle\langle ij\rangle\rangle}C_{is}^{\dagger}e^{i\tfrac{\pi}{2}\nu_{ij}}C_{js}
 + U\sum_{i,s=A,B}n_{is\uparrow}n_{is\downarrow}\;,
\end{eqnarray}
where $\langle ij\rangle$ and $\langle\langle ij\rangle\rangle$ denote the summation over all the nearest-(NN) and next-nearest-neighbor(NNN) sites, respectively;
$C_{is}^{\dagger}=(C_{is\uparrow}^{\dagger},C_{is\downarrow}^{\dagger})$ a two-component spinor with $s=A,B$ denoting two sublattices;
$\nu_{ij}=\pm 1$ the Haldane phase factor for clockwise and anticlockwise path connecting the NNN sites $i$ and $j$;
$n_{is\sigma}=C_{is\sigma}^{\dagger}C_{is\sigma}$ the electron occupation number operator at site $is$ with spin $\sigma$;
{\color{blue}$\mu$ the chemical potential to adjust the charge carrier doping $\delta$.}
The detailed lattice set-up in real and momentum space is provided in Sec. I of Sup. Mat.\cite{Zeng25-supp}.
We note that previous works for the phase diagram of HH model versus the Haldane hopping $\lambda$ and onsite electron Coulomb repulsion $U$
uncovered compensated collinear magnetism by using multiple theoretical techniques\cite{He11,Zheng15,Arun16,Wu16,Vanhala16,ifmmode16,He24,Mertz19,Mai25}.
However this compensated collinear magnetism was inaccurately classified as AFM because these works focused on studying the topological phase transition
and unveiling the novel Chern insulating state with the Chern number $C=1$\cite{Vanhala16,Mertz19,He24,Mai25},
while the electron energy dispersion in momentum space has been seldom investigated.
{\color{blue}Therefore, we investigate HH model using the cluster slave spin method[cf. Sec. IIA of Sup. Mat.\cite{Zeng25-supp}],
which has been applied to capture the crossover separating the weak- and strong-interaction regime of Hubbard model
without taking the gauge fluctuations into account\cite{WCLee17,Zeng21,Zeng22}.}
Along this way, the staggered magnetization is calculated as:
{\color{blue}
\begin{eqnarray}\label{MF-PARAM-1}
&&M = \frac{1}{4N}\sum_{is}(-1)^s[ \langle C_{is\uparrow}^{\dagger}C_{is\uparrow} \rangle -\langle C_{is\downarrow}^{\dagger}C_{is\downarrow} \rangle  ]\nonumber\\
&&\overset{\delta\to 0}{=} \frac{-1}{2N}\sum_{\bm{k}}\frac{\Delta_{\rm AFM}+\lambda(Z_{A\uparrow}+Z_{B\uparrow})\gamma_{2\bm{k}}}
{[\Delta_{\rm AFM}+\lambda(Z_{A\uparrow}+Z_{B\uparrow})\gamma_{2\bm{k}}]^2+|tZ\gamma_{\bm{k}}|^2}\;,
\end{eqnarray}
}
where $s=$ 0, 1 holds for sublattice A and B, respectively; the number of unit cells $N$, the AFM energy gap $\Delta_{\rm AFM}$,
the quasiparticle renormalization factor $Z$ and $Z_{A/B\uparrow}$, as well as the function $\gamma_{\bm{k}}$ and $\gamma_{2\bm{k}}$ respectively
from the NN hopping and the Haldane hopping are explicitly given in Sec. IIA of Sup. Mat.\cite{Zeng25-supp}.
Here, we focus on the half-filled HH model with $\mu = -U/2$ because of the particle-hole symmetry.

{\it Electronic states and phase diagram.}---To further establish the odd-parity ALM characteristics of HH model,
we first study the curves of constant energy(CCE) for up(red)- and down(blue)-spin electron quasiparticles via the electron spectrum function
$A_{\sigma}(\bm{k},\omega)=-2{\rm Im}Tr[\tilde{G}_{\sigma}(\bm{k},\omega+i\Gamma)]$ at $\lambda=0.3$ and $U=5.5$ with $\omega=$ 1 and 2
in Fig.\ref{E-EDisp-PD_GAP} (a) and (b), respectively. The electron Green's function $\tilde{G}_{\sigma}(\bm{k},\omega)$ has been provided by Eq.(18)
in Sup. Mat.\cite{Zeng25-supp}.
The major characteristics are summarized as:
(i){\color{blue}the electron spectrum function exhibits the symmetry $[C_2||C_{6z}^{1}]$ and $[C_2||\bar{E}]$,
consistent with the spin group symmetry of the hexagon-lattice system};
(ii)at low energies, the CCE's for up- and down-spin electrons are respectively centered around the Dirac point $\bm{K}$ and $\bm{K}'$ without any intersection between them,
which enables researchers to identify the odd-parity ALM via the quasiparticle scattering interference(QSI) experiment
because the QSI spectrum is closely associated with the contour of the single-particle CCE's\cite{Wang03,Zeng24,LiZeng25}.

After using the cluster slave spin method to self-consistently determine the staggered magnetization $M$
and other mean-field parameters[cf. Sec. IIA of Sup. Mat.\cite{Zeng25-supp}], as well as calculating the Thouless-Kohmoto-Nightingale-Nijs (TKNN) number,
the phase diagram versus the Haldane hopping $\lambda$ and onsite electron Coulomb repulsion $U$ is plotted in Fig.\ref{E-EDisp-PD_GAP}(c).
We found that:
(i)the complicated interaction between the Haldane hopping and the staggered magnetization
coming from the onsite electron Coulomb repulsion, as demonstrated in Eq.\eqref{MF-PARAM-1}, gives rise to a non-monotonous $\lambda$ dependence of
the critical interaction strength responsible for the transition from the paramagnetic Chern insulator(CI) state to the odd-parity ALM state;
(ii)as the staggered magnetization sets in, the breaking nonmagnetic TRS then the breaking inversion symmetry of the Bravais lattice
induced by sublattice currents transits the conventional AFM insulating(AFMI) state into the odd-parity ALM insulating(ALMI) state;
(iii)at large $\lambda$'s, in the intermediate interaction regime of grey color, the ALM insulator coexists with the Chern insulator[$C=2$],
and thus this state is dubbed as odd-parity ALM Chern insulator(ALM CI)\cite{Lin25}.
In particular, the critical interaction strength responsible for the topological phase transition
from the ALM CI state to the topologically trivial ALMI state increases monotonically with the growth of $\lambda$,
reflecting the competition between the AFM energy gap and the topological gap coming from the Haldane hopping[cf. Eq.\eqref{Gap-Function}].
We note that at $\lambda = 0$, the critical interaction strength for the transition from the paramagnetic state
to the conventional AFMI state is located around $U_{\rm AFM} = 3.1$,
which is not far from $U_{\rm AFM}\approx 3.8$ evaluated using the large-scale quantum Monte Carlo simulation\cite{Sorella_2012,Otsuka_2016,ostmeyer_2021,Raczkowski_2020,Assaad_2013,txma_2018}.
On the other hand, the accuracy of this critical interaction value
can be further improved by including more spatial fluctuations via the increment of the cluster size.

To further reveal the topological phase transition at large $\lambda's$, we investigate the up(solid)- and down(dashed)-spin single-particle energy gap,
\begin{equation}\label{Gap-Function}
\Delta_{\bm{k}\sigma}=\Delta_{\sigma}+\lambda(Z_{A\sigma}+Z_{B\sigma})\gamma_{2\bm{k}}\;,
\end{equation}
as a function of momentum along the zigzag direction in Fig.\ref{E-EDisp-PD_GAP}(d) with $\lambda=0.3$ as well as $U=$ 4.8(black), 5.1(red), 5.5(blue), and 6.1(magenta),
where $\Delta_{\sigma}$ and $\gamma_{2\bm{k}}$ are given in Eq.(15) of Sup. Mat.\cite{Zeng25-supp}.
It is found that at large $\lambda$'s, the Haldane hopping first opens opposite energy gaps at Dirac points $\bm{K}$ and $\bm{K}'$,
and drives the system into the topologically nontrivial CI state with $C=2$; As the staggered magnetization sets in
with small magnetic component $\Delta_{\sigma}$[cf. Eq.\eqref{Gap-Function}],
the sign of the single-particle energy gap at $\bm{K}$ and $\bm{K}'$ remains opposite and the system becomes an odd-parity ALM Chern insulator[$C=2$].
However, as the magnetic component $\Delta_{\sigma}$ increases and overwhelms that opened by the Haldane hopping,
energy gaps for up- and down-spin electrons are negative and positive, respectively,
and the system transits into the topologically trivial odd-parity ALMI state.

{\it Discussion and conclusion.}---We first adopt the spin group method to identify sufficient conditions
for the appearance of odd-parity ALM in collinear AFM's. It is clarified that the breaking nonmagnetic TRS,
enabling that the symmetry $[\bar{C}_2||T]$ present in the even-parity ALM no longer holds,
together with the symmetry $[C_2||\bar{E}]$ or $[C_2||M]$ that connects opposite-spin sublattices
trigger the odd-parity ALM characterized by the odd-parity spin splitting in real and momentum space.
Furthermore, on account of the breaking nonmagnetic TRS induced by sublattice currents, HH model is systematically studied using the cluster slave spin method.
Its phase diagram as a function of the Haldane hopping and onsite electron Coulomb repulsion is established,
{\color{blue}which consists of four regimes: AFMI at $\lambda=0$, CI with $C=2$, odd-parity ALM CI with $|M|>0$ and $C=2$, as well as the odd-parity ALMI with $|M|>0$ and $C=0$.}

Experimentally, the cold-atom platform using the shaking lattice technique has realized Haldane model\cite{Zheng14,Jotzu14,Das24},
while the ratio of the interaction strength $U$ versus the NN hopping $t$ can be changed flexibly
by adjusting the lattice constant of the optical lattice or the potential depth to trap the ultracold atoms,
which provides an opportunity to explore the correlated systems with topological band structure.
On the other hand, our symmetry analyses based on the spin group theory is not confined to HH model, indicating that sufficient conditions
for the appearance of odd-parity ALM can be effectively verified experimentally and theoretically\cite{Zhuang25,Luo25,Zhu26}.

\begin{acknowledgments}

Minghuan Zeng thanks Y.-J. Wang and Zhi Wang for fruitful discussions.
This work was supported by the National Key Research and Development Program of China
under Grant Nos. 2023YFA1406500 and 2021YFA1401803,
 the National Natural Science Foundation of China (NSFC)
 under Grant Nos. 12504172, 12222402, 92365101, 12474151,
 12347101, and 12274036, Beijing National
 Laboratory for Condensed Matter Physics under Grant
 No. 2024BNLCMPKF025, the Fundamental Research
 Funds for the Central Universities under Grant No.
 2024IAIS-ZX002, the Chongqing Natural Science
 Foundation under Grants Nos. CSTB2023NSCQ-JQX0024
 and CSTB2022NSCQ-MSX0568, and the Special Funding for Postdoctoral Research Projects
 in Chongqing under Grant No. 2024CQBSHTB3156.

\end{acknowledgments}

\vspace{1em}
{\it Data availability}---the data that support the findings of this study available from the corresponding
author upon reasonable request.

\bibliography{BIBHH-Model}

\end{document}